# A NOVEL APPROACH FOR WEB PAGE SET MINING

R.B.Geeta [1],  Omkar Mamillapalli [2]
Shasikumar G.Totad [3] and Prof  Prasad Reddy P.V.G.D [4]

[1]Associate Professor, Department of CSIT, GMRIT, Rajam, (A.P), India.
geetatotad@yahoo.co.in
[2]Department of CSE, GMR Information Technology, Rajam, (A.P), India.
omkar256@gmail.com
[3]Professor & HOD, Department of CSE, GMRIT, Rajam, (A.P), India.
skumartotad@yahoo.com
[4]Professor, Department of CS & SE, AndhraUniversity, Visakhapatnam,(A.P),India
Prasadreddy.vizag@gmail.com

## ABSTRACT

*The one of the most time consuming steps for association rule mining is the computation of the frequency of the occurrences of itemsets in the database.  The hash table index approach converts a transaction database to an hash index tree by scanning the transaction database only once. Whenever user requests for any Uniform Resource Locator (URL), the request entry is stored in the Log File of the server. This paper presents the hash index table structure, a general and dense structure which provides web page set extraction from Log File of server. This hash table provides information about the original database. Web Page set mining (WPs-Mine) provides a complete representation of the original database. This approach works well for both sparse and dense data distributions. Web page set mining supported by hash table index shows the performance always comparable with and often better than algorithms accessing data on flat files. Incremental update is feasible without reaccessing the original transactional database.*

## 1. Introduction

Many researchers and practitioners have been investigated Association rule mining has been for many years [1], [3], [4], [5], [6].  Agrawal  et al. introduced the problem of mining frequent itemsets for the first time[5], who proposed algorithm  Apriori. The Apriori algorithm must scan the transcation database several times and FP_growth algorithm needs to scan the database only twice. If the the database is larger, the efficiency of FP growth algorithm is  higher. To reduce scanning of database twice, the rapid association rule mining algorithm came into existence. The rapid association Rule Mining algorithm (QFP) requires to scan the transaction database once compared to FP growth algorithm, so it can increase the time efficiency of mining association rules [2].The correlations among web pages in a transactional database D can be identified using association rules. Association refers to how the web pages in the web site can be grouped. These are used to assist retail store management, marketing, grocery store problems, and inventory control. Each transaction in D is a set of web pages. Association rules are usually represented in the form A -> B, where A and B are web page sets, i.e., set of web pages. Web page sets are characterized by their frequency of occurrence in D, which is called support. Research activity usually focuses on defining efficient algorithms for web page set extraction, which represents the most computationally intensive knowledge extraction task in association rule mining [7]. In this paper, we propose a similar approach to support data mining queries. The WebPages-Mine (WPs-Mine) index is a novel data structure that provides a compact and complete representation of transactional data supporting efficient item set extraction from a relational DBMS. The following Web Pages data set shows 13 hypertext transfer protocol (http) transaction requests in given session threshold.

Table 1. Example Web Pages data set

| TID | WebPagesID | TID | WebpagesID | TID | WebPagesID |
|-----|------------|-----|------------|-----|------------|
| 1 | b,h,e,p,v,d,g | 6 | a, r, n, u ,i, b,s | 11 | r,e,h,b,a |
| 2 | m,h,n,d,b,e | 7 | b, g, h, d, e,p | 12 | z,i,a,n,r,b |
| 3 | I,f,o,h,e,c,p | 8 | a, i, b | 13 | b,d,h,p,e |
| 4 | a,w,e,k,h,j | 9 | f, e ,i, c ,h, p | | |
| 5 | d,b,e,h,n | 10 | h, a, e, b, r ,t | | |

## 2. The WPs-Mine Index

Whenever user requests for any Uniform Resource Locator, the details of request is entered into the Log File of the server. The log file entry contains various fields like IP address, time at which request is made, status code, number of bytes transferred and which page is requested. The web pages information collected in the log file is stored in the form of database. This data is stored in the form of relational model, as a relation R. Assuming some session threshold the frequency of each webpage is counted and stored in dataset as shown in table 1.

### 2.1 WPs-Hash Indexed tree Structure

The structure of the WPs-Mine index is characterized by two components: the Web Page Set-Tree and the Web Pages-Hash table tree. The two components provide two levels of indexing. The Web Pages set-Tree (WPs-Tree) is a prefix-tree which represents relation R by means of a brief and compact structure. The hash table of 26 buckets [A-Z] is created. Each bucket stores the information about the support of each web page in a assumed threshold. Each bucket holds the physical location of each web page in the website. Linked list with various nodes is attached for each bucket which holds the addresses of different web pages. The WebPages-Hash index (WPs-H index) table structure allows reading selected WPs-Tree portions during the extraction task. For each item, it stores the physical locations of all item occurrences in the WPs-Tree.

## 2.1.1 WPs-Tree:

The Web Pages set-Tree (WPs-Tree) is a prefix-tree which represents relation R by means of a short and compact structure. Implementation of the WPs-Tree is based on the FP-tree data structure, which is very effective in providing a compact and lossless representation of relation R as shown in Figure .1.

## 2.1.2 WPs-Hash-indexed tree:

The WPs-Hash-tree is a Hash table with tree structure which allows access of selected WPs-Tree portions during the extraction task. For each web page in the given website, it stores the physical locations of all web page occurrences in the Web Pages set Tree.

Figure 2 shows the WPs-Mine Hash indexed tree allows selectively accessing the WPs-Tree blocks during the extraction process. It is based on a Hash indexed Tree structure. For each item i in relation R, there is one entry in the WPs-Mine Hash indexed tree.

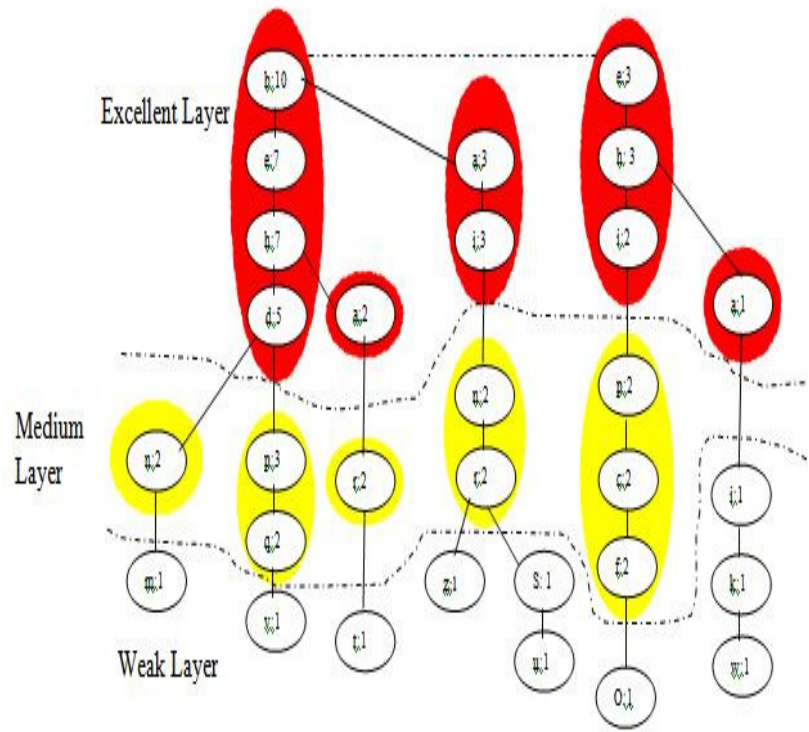

Figure 1. WPs-Mine index for the example dataset WPs-Tree

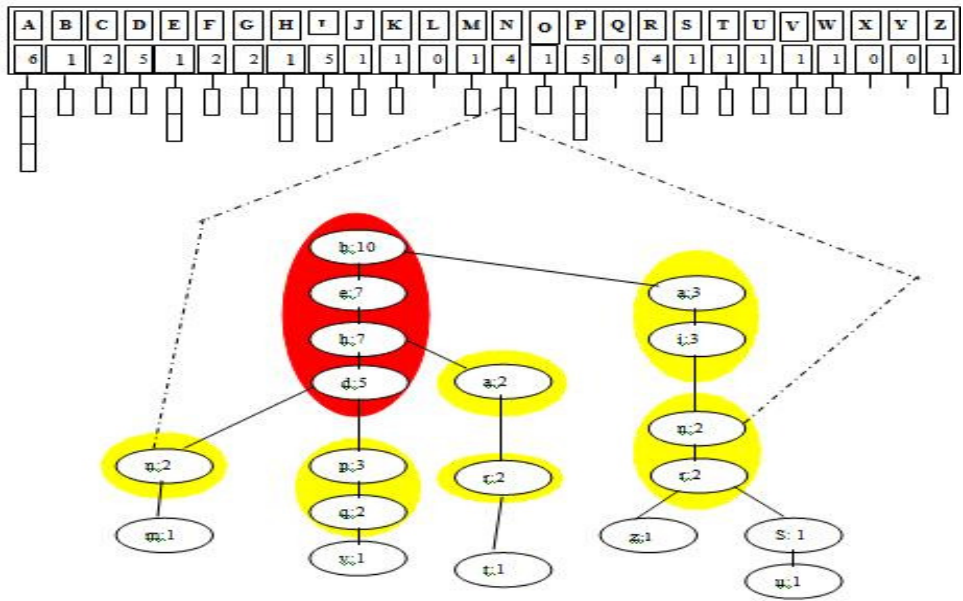

Figure 2 WPs-Mine Hash indexed tree for the example dataset WPs tree

## 2.2 WPs-Mine Data Access Methods

Three data access methods are devised to load from the WPs-Mine index the following projections of the original database: 1) Frequent WebPages-Tree to support projection-based algorithms (e.g., FP-growth [8]). 2) Support-based projection, to support level based (e.g., APRIORI [7]), and array-based (e.g., LCM v.2 [10]) algorithms. 3) web pages-based projection, to load all transactions where a specific webpage occurs, enabling constraint enforcement during the extraction process. The three access methods are described in the following sections.

### 2.2.1 Construction of Frequent WebPages-Tree:

From the relation R, the frequency of each web page is counted for a given session threshold time. The web pages are sorted in order based on its frequency but preceding in lexicographical order. In the WPs-Tree web pages are sorted by descending support lexicographical order as represented by WPs-Tree. This is represented as a prefix tree.

In the example data set, item p appears in two nodes, i.e., [p:3] and [p:2]. The access method reads two prefix paths for p, i.e., [p : 3 ->d :5 -> h : 7 !->e : 7 -> b : 10]and [p : 2 !->i : 2 !->h : 3 ->e:3] Each sub path is normalized to p node support. For example, the first prefix path, once normalized to [p:3], is [ p : 3 ->d :3 -> h : 3 -> e : 3 !->b : 3]

### 2.2.2 Support-Based division of WPs-Tree:

The support-based projection of relation R contains all transactions in R intersected with the web pages which are frequent with respect to a given support threshold (Min Sup ). The WPs-Tree paths completely represent the transactions. Web pages are sorted by decreasing support along the paths. Starting from a root node, the WPs-Tree is visited depth-first by following the node child pointer. The visit ends when a node with an Un-frequent item or a node with no children is reached.

The WPs-Tree is partitioned into three layers based on given minimum support threshold. The web pages whose support is greater than or equal to given minimum threshold is considered to be belonging to Excellent layer. The web pages whose support is greater than 1 and less than given minimum support belong to weak layer as shown in Fig. 1.

### 2.2.3 WPs-Hash-Table Tree

Log File of server contains information about how many visitors visited various web pages of web site. Given a session threshold, the frequency of each web page is counted and each page's count is stored in hash table. The hash table of 26 buckets [A-Z] is created. Each bucket stores the information about the frequency of each web page. Each bucket holds the physical location of each web page occurrences in the WPs-tree. Linked list with various nodes is attached for each bucket which holds the addresses of the occurrences of web pages in the WPs-Tree. The WebPages-Hash index (WPs-H index) table structure allows reading selected WPs-Tree portions during the extraction task. For each web page, it stores the physical locations of all page occurrences in the WPs-Tree.

### 2.3 WPs-Mine storage procedure

The organization of the WPs-Mine index is designed to minimize the cost of reading the data needed for the current extraction process. However, fetching a given record requires loading the entire disk block where the record is stored The WPs-Tree physical organization is based on the following correlation types:

**i) Intra transaction correlation:**. Web Pages appearing in a same transaction are thus intrinsically correlated. To minimize the number of read blocks, each WPs-Tree path should be partitioned into a block.

**ii) Extra transaction correlation:** In some transactions, set of web pages accessed may be same and some other pages accessed may be different, so block can be formed for common web pages accesses and separate block can be made for remaining web pages access.

### 2.3.1 WPS-Tree Layers

TheWPS-Tree is partitioned in three layers based on the node access frequency during the extraction processes. 1) the node level in the WPs-Tree, i.e., its distance from the root,2) the number of paths including it, represented by the node support, and 3) the support of its item.. The three layers are shown in Fig. 2a for the example WPs-Tree**.**

**Excellent layer**: This layer includes web pages that are very frequently accessed during the mining process. These nodes are located in the upper levels of the WPs-Tree. These web pages are most important pages as these are frequently accessed.

**Medium Layer:** This layer includes nodes that are quite frequently accessed during the mining process. This layer contains web pages which are frequently accessed during web site visits.
**Weak layer**: This layer includes the nodes corresponding to rather low support items, which are rarely accessed during the mining process. The web pages in this layer must be paid more attention to modify the content as these web pages are rarely accessed by web users.

## 3. Web Page set Mining

Web Page set mining has two sequential steps: 1) the needed index data is stored and 2) web page set extraction takes place on stored data.

### Frequent Web Pages set Extraction

This section describes how frequent web pages set extraction takes place on the WPs-Mine index. We present two approaches, denoted as FP-based and LCM-based algorithms, which are an adaptation of the FP-Growth algorithm [3][9] and LCM v.2 [10]algorithm.

### FP-based algorithm:

The FP-growth algorithm stores the data in a prefix-tree structure called FP-tree. First, it computes web page support. Then, for each transaction, it stores in the FP-tree its subset including frequent web pages. Web pages are considered one by one. For each web page, extraction takes place on the frequent-web page database, which is generated from the original FP-tree and represented in a FP-tree based structure.

## 4. EXPERIMENTAL RESULTS

The validation is done on both dense and sparse data distributions. We report the experiments on these parameters. The parameters include transaction and item cardinality, average transaction size (AvgTrSz),and data set size) as shown in Table 1. Connect is dense and medium-size data sets. Kosarak [10] is a Large and sparse data set.

### 4.1 Index Creation and Structure:

Table 2 reports both WPs-Tree and WPs-Hash index table tree size for the two data sets. The overall WPs-Mine index size is obtained by summing both contributions. The WPs-Mine indices have been created with the default value Kavg ¼ 1:2.

Furthermore, the Connect and Kosarak, and data sets have been created with KSup ¼ 0, while large synthetic data sets with KSup ¼ 0:05. In sparse data sets (e.g., Kosarak), where data are weakly correlated, data compression is low and storing the WPs-Tree requires more disk blocks.

Table 1 also shows the index creation time, which is mainly due to path correlation analysis and storage of the index paths on disk. The first factor depends on the number of WPs-Tree paths.

Table 2. Data Set Characteristics and Corresponding Indices

| Dataset | Transactions | Dataset Items | AvtrSz | Size(KB) | WPs-Tree (Records) | WPs-Hash -indexed tree (Records) | Time (sec) |
|---------|-------------|---------------|--------|----------|-------------------|----------------------------------|------------|
| CONNECT | 1200 | 38 | 7 | 8563 | 269 | 722 | 0.59 |
| KOSARAK | 1600 | 20 | 7.9 | 9564 | 291 | 729 | 0.61 |

## 4.2 Frequent Web Pages Set Extraction Performance:

The WPs-Mine structure is independent of the extraction algorithm. To validate its generality, we compared the FP-based and LCM-based algorithms with three very effective state-of-the-art algorithms accessing data on flat file.

Figure 3 generates statistics about Konark dataset containing how many WP-Tree nodes are created, how much storage required and how many WP-Tree updates. Figure 4 produces Minimum support, Number of records, Number of columns and WP-Tree updates. Figure 5 shows what is the storage required by WP-Hash-index Tree along with generation time for the tree. Similarly Figure 6 generates statistics about Connect Dataset.

Figure 8 compares the FP-based algorithm with the FP-growth algorithms [3] on flat file, all characterized by a similar extraction approach. For real data sets (Connect, and Kosarak), differences in CPU time between the FP-based and the Prefix-Tree algorithms are not visible for high supports, while for low supports the FP-based approach always performs better than Prefix-Tree. Comparison is made between connect and Kosarak dataset. In connect data set WPs-Hash index Tree 3 times faster than WPs-tree.so WPs-Hash index Tree reduced the search time and I/O cost. In Kosarak data set WPs-Hash index Tree 3 times faster than WPs-tree.so WPs-Hash index Tree reduced the search time and I/O cost. This effect is particularly relevant for low supports, because representing in memory a large portion of the data set may significantly reduce the space for the extraction task, hence causing more memory swaps. Figure 9 displays run Time Comparison Between (a)Connect and (b) Kosarak dataset

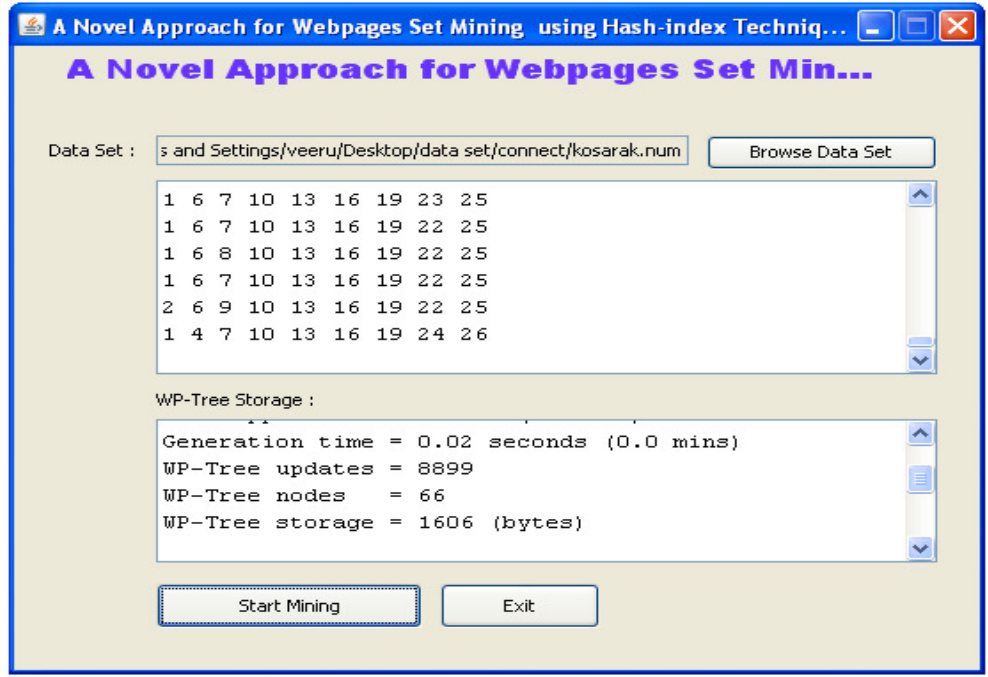

Figure 3. Kosark statistics

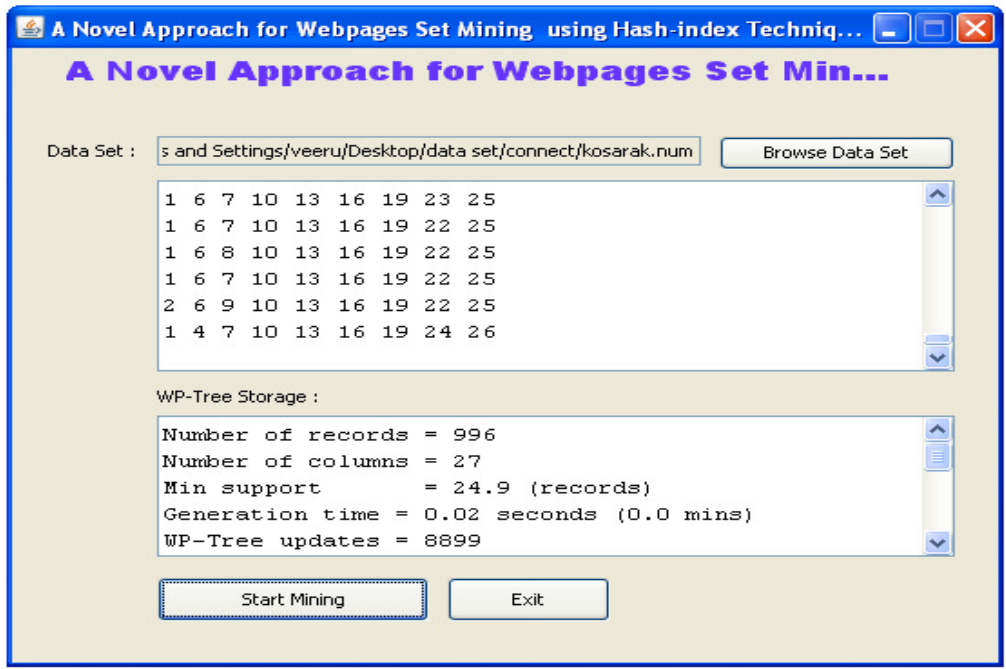

Figure 4. The Dataset Kosark statistics

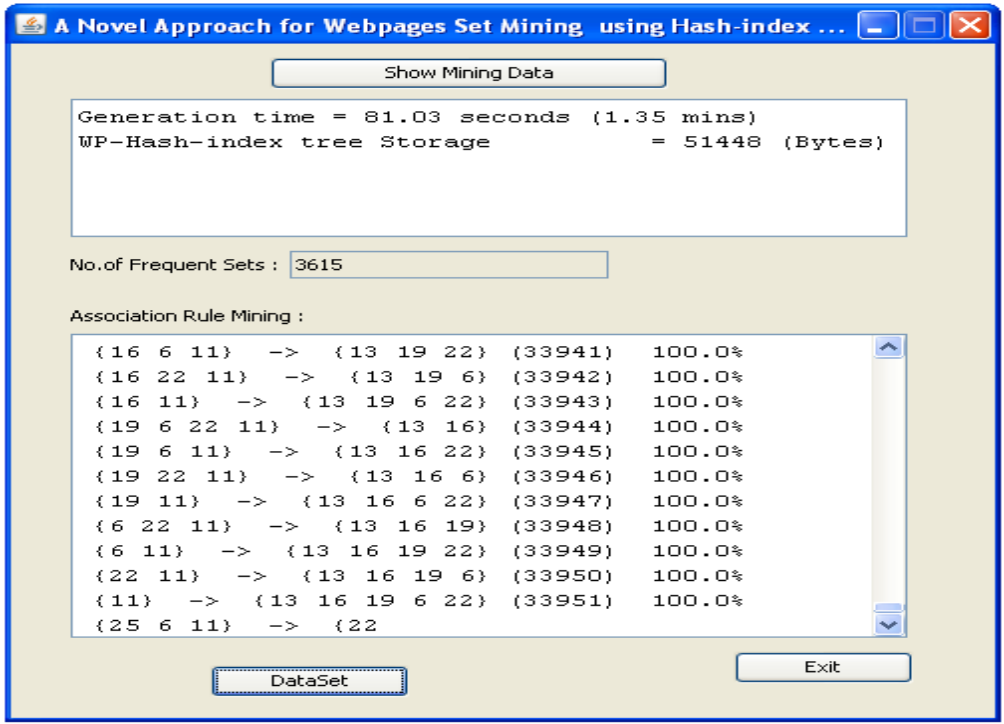

Figure 5. The Database Kosark statistics

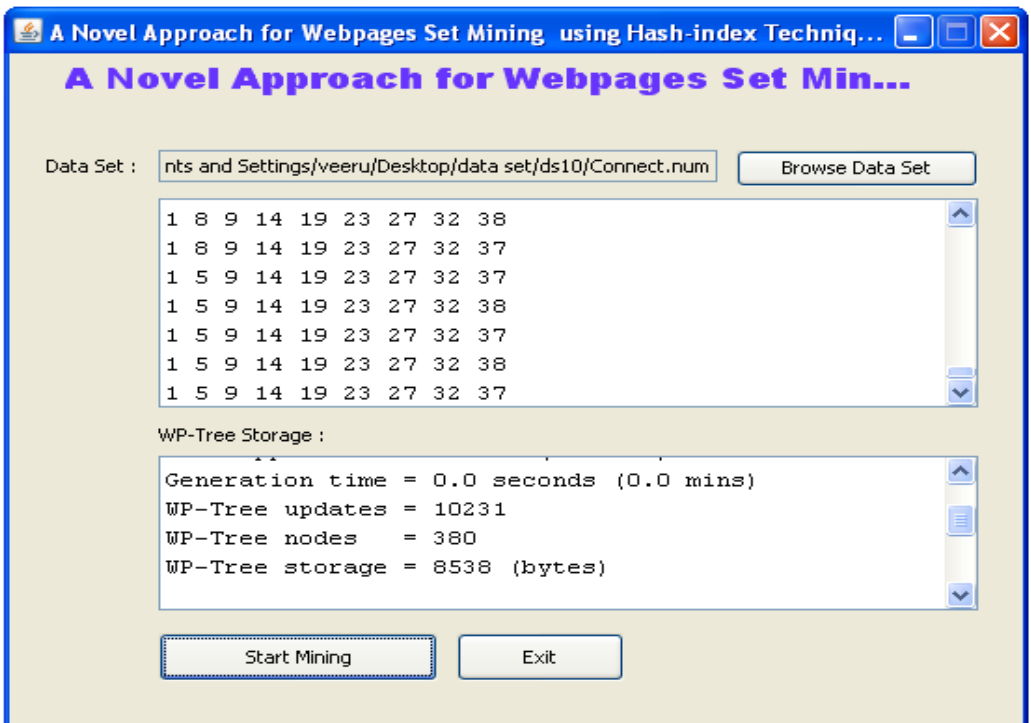

Figure 6. The Dataset Connect statistics

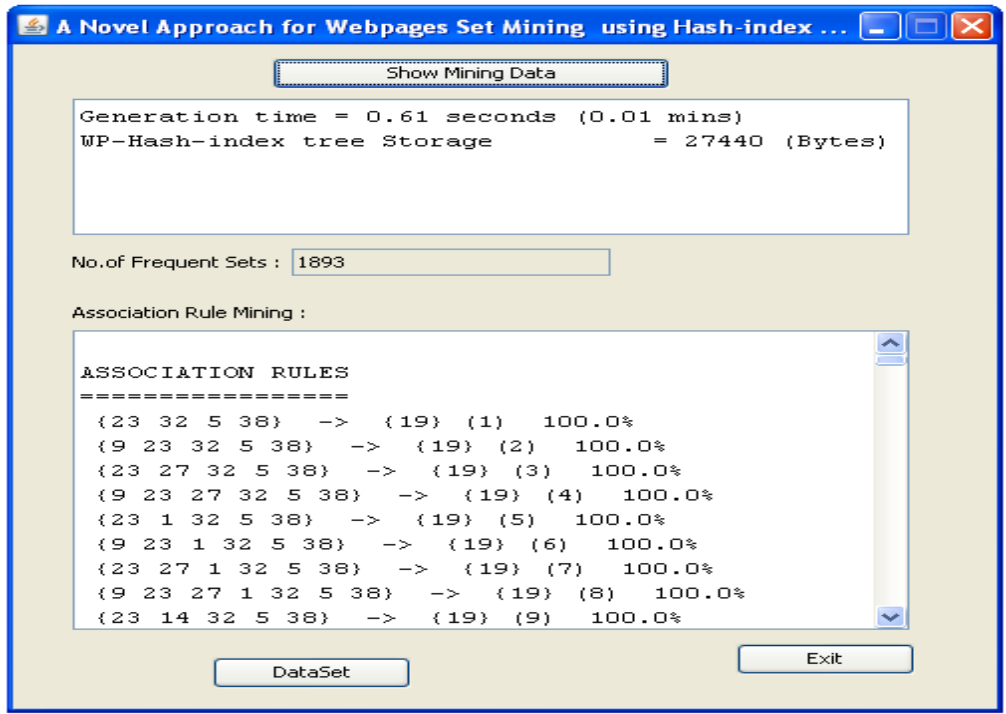

Figure 7. The Dataset  Connect  statistics

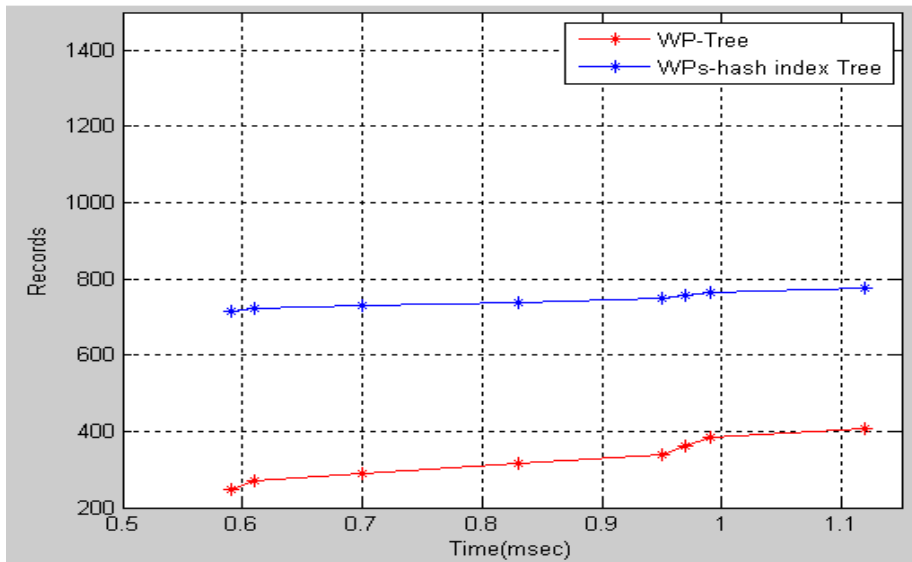

Figure 8.  Frequent WPs set extraction time for the FP-based algorithm. (a) Connect

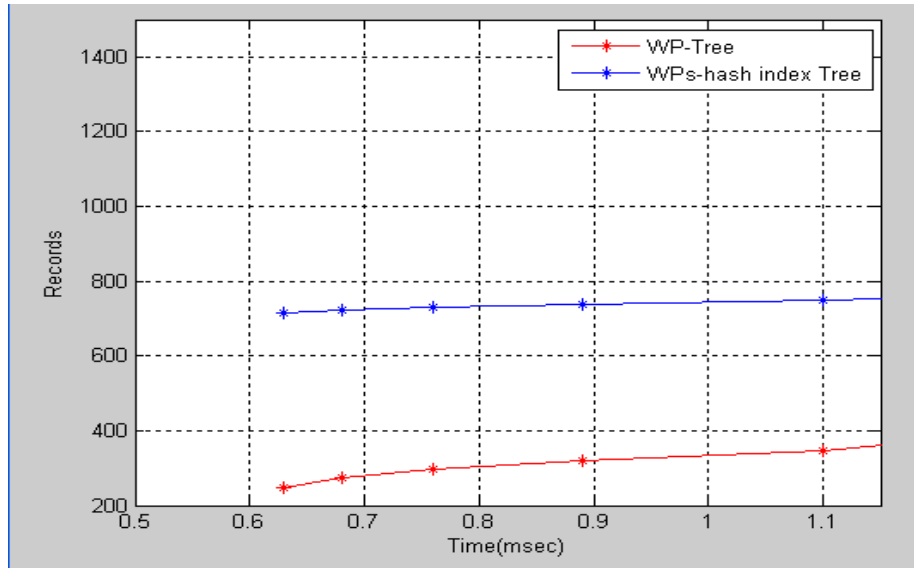

Figure .8 Frequent WPs set extraction time for the FP-based algorithm. (b) Kosarak.

## 4.3 Comparison

Run Time Comparison between WP-Tree and WPs-Hash index Tree for Dataset Connect and Kosarak dataset is shown in Figure 9a & Figure 9 b respectively. . In connect data set WPs-Hash index Tree 3 times faster than WPs Tree. So WPs-Hash index Tree reduced the search time and I/O cost. In Kosarak data set WPs-Hash index Tree 3 times faster than WPs-tree.so WPs-Hash index Tree reduced the search time and I/O cost.

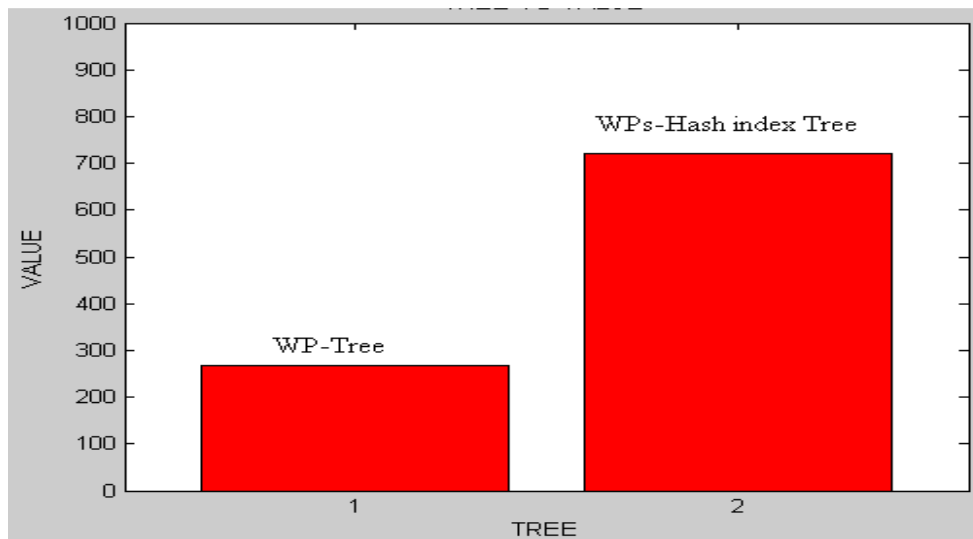

Figure 9 a. Run Time Comparison Between WPs-Tree and WPs-Hash index Tree Connect dataset

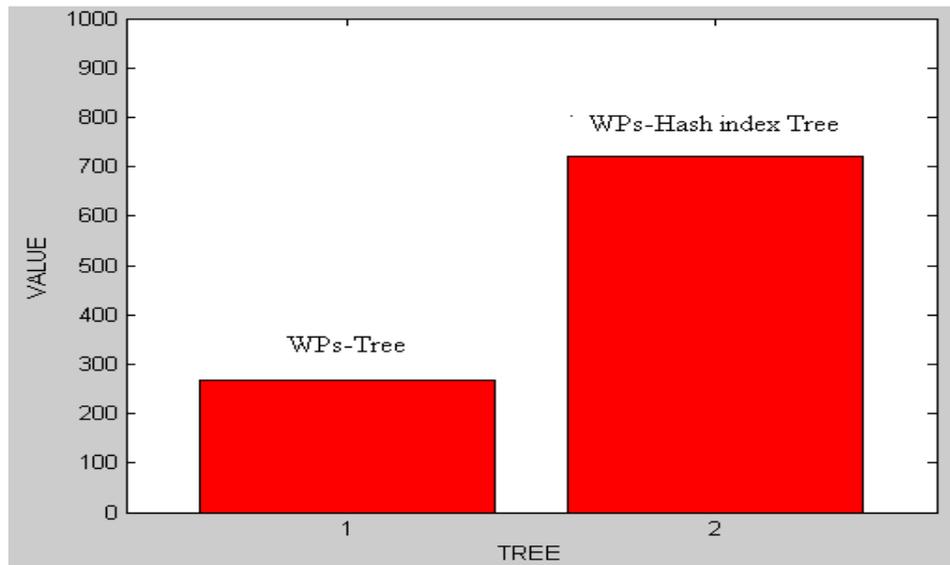

Figure 9 b. Run Time Comparison Between WPs-Tree and WPs-Hash index Tree
Kosarak dataset

## 5. CONCLUSION

Efficient Data mining algorithms play an important role when transaction databases are very large. Since transaction databases are huge, it may not be stored in main memory. This Hash-indexing approach helps in completion of mining process through scanning the transaction database only once. This experiment showed that this algorithm is more efficient compared to other existing algorithms. The WPs-Hash index tree structure provides efficient access reducing I/O time. Further extension may be to have still efficient compact structure for different data distributions and incremental updating of index. Incremental update is feasible without re-accessing the original transactional database. Incremental update can be done considering the transaction where session was ended.